\documentclass[aps,preprint,nofootinbib]{revtex4}%
\usepackage{amsfonts}
\usepackage{amsmath}
\usepackage{amssymb}
\usepackage[english]{babel}
\usepackage{graphicx}
\usepackage{subfig}
\usepackage{float}
\usepackage{graphicx}%
\providecommand{\U}[1]{\protect\rule{.1in}{.1in}}

\begin{document}
\title{Charged black strings and black branes in Lovelock theories}
\author{Alex Giacomini$^{1}$, Marcela Lagos$^{2}$, Julio Oliva$^{2}$, Aldo Vera$^{2}$}
\affiliation{$^{1}$Instituto de Ciencias F\'{\i}sicas y Matem\'{a}ticas, Universidad
Austral de Chile, Casilla 567, Valdivia, Chile}
\affiliation{$^{2}$Departamento de F\'{\i}sica, Universidad de Concepci\'{o}n, Casilla
160-C, Concepci\'{o}n, Chile}

\begin{abstract}
It is well known that the Reissner-Norstrom solution of Einstein-Maxwell
theory cannot be cylindrically extended to a higher dimension, as with the black hole solutions in vacuum. In this paper we show that this result is
circumvented in Lovelock gravity. We prove that the theory containing only the
quadratic Lovelock term, the Gauss-Bonnet term, minimally coupled to a $U(1)$
field, admits homogeneous black string and black brane solutions characterized
by the mass, charge and volume of the flat directions. We also show that
theories containing a single Lovelock term of order $n$ in the Lagrangian
coupled to a $(p-1)$-form field admit simple oxidations only when $n$ equals
$p$, giving rise to new, exact, charged black branes in higher curvature gravity. For
general relativity this stands for a Lagrangian containing the Einstein-Hilbert
term coupled to a massless scalar field, and no-hair theorems in this case
forbid the existence of black branes. In all these cases the field equations
acquire an invariance under a global scaling scale transformation of the metric. As explicit
examples we construct new magnetically charged black branes for cubic Lovelock
theory coupled to a Kalb-Ramond field in dimensions $(3m+2)+q$, with $m$ and
$q$ integers, and the latter denoting the number of extended flat directions. We
also construct dyonic solutions in quartic Lovelock theory in dimension
$(4m+2)+q$.

\end{abstract}
\maketitle

\section{Introduction}

In general relativity (GR) in vacuum, black branes are objects that are
intrinsic to dimensions higher than four. Since the direct product of Ricci
flat manifolds is still Ricci flat, one can construct black branes in GR by a
simple oxidation of the black hole solutions of a given dimension $d$ to
dimension $D=d+q$ by adding $q$ flat directions. In spite of their simple
origin, these spacetimes possess very rich physics. They describe the rapidly
rotating limits of the black ring in five dimensions \cite{ER} as well as the
Myers-Perry black hole in dimension greater than or equal to six \cite{MP}, since
in the latter case the angular momentum is not bounded from above. These
rapidly rotating objects inherit the instability of the black branes, the
Gregory-Laflamme instability, which is triggered by perturbations traveling
along the extended direction with wavelength above a critical value
\cite{GL}-\cite{GLCharged}. Therefore, one can in principle get rid of the
instability by the compactification of the extra direction with a radius below
the critical wavelength, but then, decreasing the mass of the black hole on
the brane, and therefore its radius, the instability reappears. These results
were extended to charged black strings in low energy string gravity with
isometry $\mathrm{I\!R}^{10-d}\times\mathrm{I\!R}_{t}\times S^{d-2}$ in
dimension ten, where a dilaton and a $(d-3)$-form [with field strength a
$(d-2)$-form] are present. Evidence on the stability of the extremal cases was
reported also in \cite{GLEvidence}. In these solutions it is important to
notice that the dilaton cannot be turned off without turning off the
$(d-3)$-form field, and therefore, in the case $d=4$ the solution is not
continuously connected to an oxidation of the Reissner-Norstrom solution. When
the instability takes place, numerical simulations in five dimensions show
that the black string may brake into a chain of black holes \cite{LP1}, \cite{LP2}, \cite{LP3}, in a finite value of the proper time of an asymptotic
observer for generic initial data, providing evidence on a violation of cosmic
censorship in asymptotically Kaluza-Klein spacetimes. Recent simulations show
a similar behavior for the late time evolution of the black ring instability
\cite{FIGUERAS1} as well as in the ultraspinning regime of the Myers-Perry
solution in dimension six \cite{FIGUERAS2}, providing evidence on violations
of cosmic censorship, for generic initial data as well, in an asymptotically
flat spacetime in higher dimensions. For any theory predicting corrections to
GR, in particular in the context of string theory, one expects that as the
perturbed black string evolves it may unveil regions with high curvature and
therefore higher curvature corrections may play a role on the evolution of the
spacetime. Consequently, it is natural to look for black brane solutions in
the presence of higher curvature terms.

A good model for exploring the physics of higher curvature gravity are
Lovelock theories. These are the most general theories having only the metric
as a fundamental field, with second order field equations \cite{LOVELOCK}.
Even more, the quadratic Gauss-Bonnet theory appears in the low energy limit
of the bosonic and heterotic string theories. The Lagrangian of Lovelock
theory is constructed by linearly combining the dimensional continuations of
all the Euler densities of even dimensions lower than $D$, each term being a
homogeneous polynomial of order $l$ on the Riemann tensor, and is given by
\bigskip%
\begin{equation}\label{Lagfull}
\mathcal{L}=\sqrt{-g}\sum_{l=0}^{\bar{l}}c_{l}\mathcal{L}^{(l)}\ ,
\end{equation}
with
\begin{equation}
\label{eq:Ln}\mathcal{L}^{(l)}=\frac{1}{2^{l}}\delta^{C_{1} D_{1}\cdots C_{l}
D_{l}}_{A_{1} B_{1}\cdots A_{l} B_{l}}R^{A_{1}B_{1}}_{C_{1}D_{1}}\cdots
R^{A_{l}B_{l}}_{C_{l}D_{l}}.
\end{equation}
Here $\bar{l}\leq\left[  \frac{D-1}{2} \right]  $. The $l=0$ contribution is a
bared cosmological term, the $l=1$ is the Einstein-Hilbert Lagrangian and the
$l=2$ term is the Gauss-Bonnet term. Explicitly,
\begin{align}
\mathcal{L}^{(0)}  &  =1\ ,\\
\mathcal{L}^{(1)}  &  =\frac{1}{2}\delta^{CD}_{AB}R^{AB}_{CD}=R\ ,\\
\mathcal{L}^{(2)}  &  =\frac{1}{4}\delta^{C_{1}D_{1}C_{2}D_{2}}_{A_{1}B_{1}%
A_{2}B_{2}}R^{A_{1}B_{1}}_{C_{1}D_{1}}R^{A_{2}B_{2}}_{C_{2}D_{2}}=R^{2}-4
R_{AB}R^{AB}+R_{ABCD}R^{ABCD} \ ,\label{L2}\\
\mathcal{L}^{3}  &  =\frac{1}{8}\delta^{C_{1}D_{1}C_{2}D_{2}C_{3}D_{3}}%
_{A_{1}B_{1}A_{2}B_{2}A_{3}B_{3}}R^{A_{1}B_{1}}_{C_{1}D_{1}}R^{A_{2}B_{2}%
}_{C_{2}D_{2}}R^{A_{3}B_{3}}_{C_{3}D_{3}}\\
&  =R^{3}-12 R R_{AB}R^{AB}+16 R_{AB}R^{A}{}_{C}R^{BC}+24R_{AB}R_{CD}%
R^{ABCD}+3R R_{ABCD}R^{ABCD}\nonumber\\
&  -24R_{AB}R^{A}{}_{CDE}R^{BCDE}+4R_{ABCD}R^{ABEF}R^{CD}{}_{EF}-8R_{ABC}%
{}^{D} R^{AECF}R^{B}{}_{EDF}\ .
\end{align}

\bigskip

This theory shares many of the properties of GR. For generic values of the
dimensionful couplings $c_{l}$, the theory admits a Birkhoff's theorem in the
sense that spherical symmetry implies the existence of an extra Killing vector
which is timelike outside the event horizon of the black hole solutions \cite{ZEGERS}, \cite{DESERREDUX}. The lapse function of the black hole
solution is determined up to a polynomial equation (Wheeler's polynomial) \cite{BD}, \cite{WHEELER1}, \cite{WHEELER2}, and each of the solutions of the
polynomial has an asymptotic behavior that matches one of the possible
maximally symmetric vacua of the theory \cite{GG}, \cite{CHARMOUSISREVIEW}. As
in GR one would like to construct black branes, nevertheless, for generic
values of the couplings, it is not possible to trivially oxidate the black
hole solution of a given dimension $d$ and one has to rely on numerical
methods to construct homogeneous black strings. This is done by warping the
extended directions with a function of the radial coordinate \cite{BSEGB1},
\cite{BSEGB2}, \cite{BSEGB3}, which in the Kaluza-Klein language is read as a
dilaton with a nontrivial radial profile. Notwithstanding, when all but one
of the constants $c_{l}$ is nonvanishing the theory contains black holes \cite{BHS}, \cite{BHSTOP} (see also \cite{Cai}), and mimics the behavior of
general relativity since in this case the black holes in vacuum can be
trivially oxidated \cite{GOT} \footnote{This result can be extended to
Lovelock theories with a unique maximally symmetric solution provided the
black hole on the transverse section is warped by a function of the extended
direction \cite{GOT}, \cite{KASTORMANN}. Also, it has been recently shown that for GR with a negative cosmological constant, homogenous black strings can be constructed at the cost of introducing a massless scalar field which is linear along the extended direction \cite{CisternaOliva}.}. The black holes on the transverse
section can also be obtained from the general solution with generic couplings
in the regime where the radial coordinate is small compared with all the
length scales involved in the theory, i.e. $r\ll|c_{l}/c_{1}|^{\frac{1}{2l-2}}$ for
all $l>1$. This is natural since the highest curvature term must dominate the
dynamics in the high curvature regime. Recently it was shown that the black
strings constructed in this manner also suffer from a Gregory-Laflamme
instability, under spherically symmetric perturbations that travel along the
extended directions, in the quadratic and cubic Lovelock theories \cite{US1},
\cite{US2}, respectively (see also \cite{USProceeding} for black branes).

The black hole solutions can be charged under a Maxwell field \cite{BHS} and
it is therefore natural to explore whether these spacetimes may lead to charged
black branes. Surprisingly enough, in this paper we show that the solutions of
the Gauss-Bonnet theory minimally coupled to a Maxwell field can be trivially
oxidated. In GR the equations along the extended directions behave as scalars
under the change of coordinates on the brane, and turn out to be incompatible with
the trace of the field equations on the brane. Nevertheless, here we show that
these equations are indeed compatible if the gravitational theory contains
only the Gauss-Bonnet term, i.e. $c_{l}=c_{2}\delta_{(l,2)}$ in Eq. (\ref{Lagfull}). This leads to
new black strings and black branes in Gauss-Bonnet-Maxwell theory, which is
shown in Sec. II. In Sec. III we show an interesting extension of this
result by proving that if the action contains a single Lovelock term of order
$n$ plus the Maxwell action for a $(p-1)$-form field (with field strength
$F_{(p)}$), one can construct homogeneous strings and branes with
$F_{(p)}\wedge\star F_{(p)}\neq0$ only if $n=p$. This implies that the
solutions of GR coupled to a massless scalar field can be trivially embedded
in the same theory in higher dimensions by adding flat directions,
nevertheless no-hair theorems imply that no black branes can be constructed in
this manner, and that the only static solution with spherical symmetry on the
brane is the Janis-Newman-Winicour naked singularity in arbitrary dimension
$d$ \cite{JNW}, \cite{Xanthopoulos:1989kb}, times $\mathrm{I\!R}^{q}$. In
Sec. IV we construct new solutions for the cubic Lovelock theory coupled to a
three-form field strength, which exploiting the results of Sec. III can be oxidated to describe magnetically charged black branes in
dimensions $(3m+2)+q$, with $m$ and $q$ integers, for which the horizon is
given by a product of $m$ three-dimensional spaces of constant curvature
$\gamma$ times $\mathrm{I\!R}^{q}$. We also construct new magnetic black brane
solutions for the quartic Lovelock theory $\mathcal{R}^{4}$ coupled to a
three-form fundamental field, in dimensions $(4m+2)+q$, as well as dyonic
solutions in dimension $(2m+2)+q$.

\section{New charged black strings in Gauss-Bonnet-Maxwell theory}

Let us first review the obstruction on the construction of black branes in
Einstein-Maxwell theory by cylindrical oxidation of the Reissner-Norstrom
solution. The field equations for Einstein-Maxwell theory are%
\begin{equation}
\mathcal{E}^{\left(  1,2\right)  }:=R_{AB}-\frac{1}{2}g_{AB}R-\left(
F_{AC}F_{B}^{\ C}-\frac{1}{4}g_{AB}F_{CD}F^{CD}\right) \, =\,0\, .
\end{equation}
Assuming an electric ansatz of the form%
\begin{align}
ds^{2}  &  =d\tilde{s}_{D-q}^{2}+d\vec{x}_{q}^{2}=-f\left(  r\right)
dt^{2}+\frac{dr^{2}}{g\left(  r\right)  }+r^{2}d\Omega_{D-2-q}^{2}+d\vec
{x}_{q}^{2}\ ,\label{ansatzmetric}\\
A  &  =A_{t}\left(  r\right)  dt\ , \label{ansatzelectric}%
\end{align}
one obtains two sets of equations:%
\begin{equation}
\tilde{R}_{\mu\nu}-\frac{1}{2}\tilde{g}_{\mu\nu}\tilde{R}=\tilde{F}_{\mu
\alpha}\tilde{F}_{\nu}^{\ \alpha}-\frac{1}{4}\tilde{g}_{\mu\nu}\tilde
{F}_{\alpha\beta}\tilde{F}^{\alpha\beta}\ , \label{enomega}%
\end{equation}
and%
\begin{equation}
-\frac{1}{2}g_{x_{i}x_{i}}\tilde{R}=-\frac{1}{4}g_{x_{i}x_{i}}\tilde
{F}_{\alpha\beta}\tilde{F}^{\alpha\beta}\ \text{(no sum over }x_{i}\text{)},
\label{extended}%
\end{equation}
where the Greek indices run along the $d=(D-q)$-dimensional manifold with line
element $d\tilde{s}^{2}$, objects with a tilde are intrinsically defined in
this manifold and $x_{i}$, with $i=1,...,q$, stand for the Cartesian
coordinates along the extended, flat directions. Equation (\ref{enomega})
leads to the Reissner-Norstrom solution in $D-q$ dimensions with the gauge
potential given by $A_{t}=-Q_{e}/r^{d-3}+A_{\infty}$. Inserting the trace of
(\ref{enomega}) on Eq. (\ref{extended}) one finds that%
\begin{equation}
\left(  D-q-2\right)  \tilde{F}^{2}=\left(  D-q-4\right)  \tilde{F}^{2}\ ,
\label{inconsistentgrmax}%
\end{equation}
which is compatible only if the electric charge vanishes, and therefore
$A_{t}$ is constant. As expected, this shows that the Reissner-Norstrom solution
cannot be extended cylindrically to higher dimensions by just adding flat
directions. The factor of the left-hand side of (\ref{inconsistentgrmax}) is
determined by the fact that
\begin{equation}
\tilde{g}^{\mu\nu}\tilde{G}_{\mu\nu}=\frac{2-(D-q)}{2}\tilde{R}\ .
\end{equation}

\bigskip

For the Gauss-Bonnet theory, whose Lagrangian is given by $\mathcal{L}%
^{\left(  2\right)  }$ in Eq. (\ref{L2}) the generalized Einstein tensor
reads%
\begin{equation}
\mathcal{G}_{AB}^{\left(  2\right)  }=2RR_{AB}-4R_{AC}R^{C}_{B}-4R_{CD}%
R^{CD}_{AB}+2R_{ACDE}R_{B}^{CDE}-\frac{1}{2}g_{AB}\mathcal{L}^{(2)}\ ,
\end{equation}
and one can see that in dimension $D$%
\begin{equation}
g^{AB}\mathcal{G}_{AB}^{\left(  2\right)  }=\frac{4-D}{2}\mathcal{L}^{\left(
2\right)  }\ .
\end{equation}
We can couple $\mathcal{L}^{\left(  2\right)  }$ to Maxwell's theory and the
field equations then read%
\begin{equation}
\mathcal{E}^{\left(  2,2\right)  }:=\mathcal{G}_{AB}^{\left(  2\right)
}-\left(  F_{AC}F_{B}^{\ C}-\frac{1}{4}g_{AB}F_{CD}F^{CD}\right)  =0\ .
\label{EGBMax}%
\end{equation}
Splitting the equations using the ansatz of the electrically charged black
brane (\ref{ansatzmetric}) and (\ref{ansatzelectric}) one obtains%
\begin{equation}
\mathcal{\tilde{G}}_{\mu\nu}^{\left(  2\right)  }=\tilde{F}_{\mu\alpha}%
\tilde{F}_{\nu}^{\ \alpha}-\frac{1}{4}\tilde{g}_{\mu\nu}\tilde{F}_{\alpha
\beta}\tilde{F}^{\alpha\beta}\ , \label{onomegaGB}%
\end{equation}
and%
\begin{equation}
-\frac{1}{2}g_{x_{i}x_{i}}\mathcal{\tilde{L}}^{\left(  2\right)  }=-\frac
{1}{4}g_{x_{i}x_{i}}\tilde{F}_{\alpha\beta}\tilde{F}^{\alpha\beta}\ \text{(no
sum over }x_{i}\text{)\ }. \label{extendedGB}%
\end{equation}
Now, inserting the trace of the equation defined on the $(D-q)$-dimensional
manifold (\ref{onomegaGB}) into (\ref{extendedGB}) one obtains%
\begin{equation}
\left(  D-q-4\right)  \tilde{F}^{2}=\left(  D-q-4\right)  \tilde{F}^{2}\ ,
\end{equation}
which is trivially fulfilled and imposes no extra constraint on the electric
field. Therefore, one is left with the system of equations projected on the
brane (\ref{onomegaGB}), which admits black hole solutions with charge and
mass \cite{BHS}. This shows that it is enough to solve the system
(\ref{onomegaGB}) which will provide a solution of Gauss-Bonnet-Maxwell theory
in $D-q$ dimensions, and admits a cylindrical uplift to dimension $D$ with planar
extra directions. Explicitly, the following brane spacetime is a solution of
the Gauss-Bonnet-Maxwell theory (\ref{EGBMax}):%
\begin{align}
ds^{2}  &  =-\left(  1-\left(  \frac{\mu}{r^{D-q-5}}-\frac{Q^{2}}{r^{2\left(
D-q-4\right)  }}\right)  ^{\frac{1}{2}}\right)  dt^{2}+\frac{dr^{2}}{1-\left(
\frac{\mu}{r^{D-q-5}}-\frac{Q^{2}}{r^{2\left(  D-q-4\right)  }}\right)
^{\frac{1}{2}}}+r^{2}d\Omega_{D-q-2}^{2}+d\vec{x}_{q}^{2}\ , \\
A  &  =-\sqrt{(D-q-4)(D-q-2)}\frac{Q}{r^{D-q-3}}dt\ ,
\end{align}
where $\mu$ and $Q$ are integration constants and $d\Omega_{D-q-2}$ is the
line element of a $(D-q-2)$-sphere. This spacetime describes a black brane for
$D-q>5$. One can check the consistency with the general argument provided
above, and show explicitly that the field equations along the extended
directions $\vec{x}$ are trivially fulfilled.

\bigskip

It is interesting to notice that the dynamics of this theory is invariant
under global scale transformations of the metric, since under $g_{AB}\rightarrow\xi
^{-1}g_{AB}$, the field equations (\ref{EGBMax}) scale as%
\begin{equation}
\mathcal{E}^{\left(  2,2\right)  }_{AB}=0\rightarrow\xi\mathcal{E}^{\left(
2,2\right)  }_{AB}=0\ .
\end{equation}
This is not a symmetry of the action but it is a symmetry of the field equations.

\section{Lovelock theories with a single term and $p$-forms}

The results of the previous section immediately motivate the exploration of
Lovelock theories containing a single Lovelock term of order $n$, coupled to a
$p$-form field strength, with field equations given by%
\begin{equation}
\mathcal{E}^{\left(  n,p\right)  }:=\mathcal{G}_{AB}^{\left(  n\right)
}-\left(  F_{AC_{1}\cdots C_{p-1}}F_{B}^{\ C_{1}\cdots C_{p-1}}-\frac{1}{2p}%
g_{AB}F_{C_{1}\cdots C_{p}}F^{C_{1}\cdots C_{p}}\right)  =0\ , \label{np}%
\end{equation}
and the Maxwell equation for a $\left(  p-1\right)  $-form%
\begin{equation}
\nabla_{A}F^{AC_{1}\cdots C_{p-1}}=\, 0\ .
\end{equation}
The term in the parentheses of (\ref{np}) corresponds to the energy-momentum
tensor of a $(p-1)$-form field with field strength $F_{\left(  p\right)  }$,
and $\mathcal{G}_{AB}^{\left(  n\right)  }$ is the Lovelock tensor of order $n$, i.e.%
\begin{equation}
\mathcal{G}_{\ B}^{A\left(  n\right)  }:=-\frac{1}{2^{n+1}}\delta
_{BD_{1}\cdots D_{2n}}^{AC_{1}\cdots C_{2n}}R_{\ \ C_{1}C_{2}}^{D_{1}D_{2}%
}\cdots R_{\ \ C_{2n-1}C_{2n}}^{D_{2n-1}D_{2n}}\ ,
\end{equation}
which reduces to the Einstein tensor when $n=1$. This tensor is obtained by
the Euler-Lagrange derivative with respect to the metric of the integral spacetime integral of $\sqrt{-g}\mathcal{L}%
^{\left(  n\right)  }$ with $\mathcal{L}^{\left(  n\right)  }$ defined in
(\ref{eq:Ln}), and it is therefore symmetric and divergenceless. It is important
to notice the following identity:
\begin{equation}
g^{AB}\mathcal{G}_{AB}^{\left(  n\right)  }=\left(  \frac{2n-D}{2}\right)
\mathcal{L}^{\left(  n\right)  }\ .
\end{equation}

We are interested in homogenous black strings and branes with metrics of the
form (\ref{ansatzmetric}), and therefore, as in the electromagnetic case, let
us assume that the fundamental $(p-1)$-form field depends only on the
coordinates on the brane. This requirement is general enough also to admit
solutions with magnetic charges. We can now project the equations of the
Lovelock theory supported by a $(p-1)$-form (\ref{np}) on the brane and along
the $q$ extended directions, which respectively lead to%
\begin{align}
\mathcal{\tilde{G}}_{\mu\nu}^{\left(  n\right)  }  &  =\left(  \tilde{F}%
_{\mu\alpha_{1}\cdots \alpha_{p-1}}\tilde{F}_{\nu}^{\ \alpha_{1}\cdots \alpha_{p-1}%
}-\frac{1}{2p}\tilde{g}_{\mu\nu}\tilde{F}_{\alpha_{1}\cdots \alpha_{p}}\tilde
{F}^{\alpha_{1}\cdots \alpha_{p}}\right)  \ ,\label{onomeganp}\\
-\frac{1}{2}g_{x_{i}x_{i}}\mathcal{\tilde{L}}^{\left(  n\right)  }  &
=-\frac{1}{2p}g_{x_{i}x_{i}}\tilde{F}_{\alpha_{1}\cdots\alpha_{p}}\tilde
{F}^{\alpha_{1}\cdots\alpha_{p}}\ \text{(no sum over }x_{i}\text{)}.
\label{extendednp}%
\end{align}
Inserting (\ref{extendednp}) on the trace of the $(D-q)$-dimensional equations
(\ref{onomeganp}), one obtains%
\begin{equation}
\left(  D-q-2n\right)  \tilde{F}_{\alpha_{1}\cdots\alpha_{p}}\tilde{F}%
^{\alpha_{1}\cdots\alpha_{p}}=\left(  D-q-2p\right)  \tilde{F}_{\alpha
_{1}\cdots\alpha_{p}}\tilde{F}^{\alpha_{1}\cdots\alpha_{p}}\ ,
\end{equation}
which for $\tilde{F}_{\alpha_{1}\cdots\alpha_{p}}\tilde{F}^{\alpha_{1}%
\cdots\alpha_{p}}\neq0$ is consistent only if $n=p$. This shows that we can
cylindrically uplift the solutions of Lovelock theory with a single Lovelock term
of order $n$, only when they are supported by an $(n-1)$-form field. It is
therefore enough to solve the equations on the brane (\ref{onomeganp}) to find
new extended black objects. The field equations of such theory also acquire a
global scaling symmetry since under$\ g_{AB}\rightarrow\xi^{-1}g_{AB}$, one
has%
\begin{equation}
\mathcal{E}_{AB}^{\left(  n,n\right)  }=0\rightarrow\xi^{n-1}\mathcal{E}%
_{AB}^{\left(  n,n\right)  }=0\ .
\end{equation}
Note that in order to have a nonvanishing contribution of the Lovelock theory
of order $n$ to the field equations we need $d>2n$.

\bigskip

It is interesting to notice that Einstein theory supported by a single
massless scalar field belongs to the special class of theories singled out by
the previous arguments. Indeed in such a case the equations on the brane read%
\begin{equation}
\tilde{R}_{\mu\nu}-\frac{1}{2}\tilde{g}_{\mu\nu}\tilde{R}=\partial_{\mu}%
\phi\partial_{\nu}\phi-\frac{1}{2}\tilde{g}_{\mu\nu}\tilde{g}^{\alpha\beta
}\partial_{\alpha}\phi\partial_{\beta}\phi\ , \label{GRSCALARonomega}%
\end{equation}
while the equations along the extended directions reduce to%
\begin{equation}
\tilde{R}=\tilde{g}^{\alpha\beta}\partial_{\alpha}\phi\partial_{\beta}\phi\ ,
\end{equation}
which is trivially implied by the trace of (\ref{GRSCALARonomega}). Due to
no-hair theorems, it is not possible to obtain cylindrically extended black objects in this
system, nevertheless for spherical symmetry on the brane, this shows that one can indeed cylindrically extend the singular Janis-Newman-Winicour \cite{JNW} solution from four
dimensions to $D=(4+q)$-dimensions as a solution of the Einstein-Klein-Gordon
system in dimension $D$. This can be extended to arbitrary $d$ by considering
the Zannias-Xanthopoulos singular solution \cite{Xanthopoulos:1989kb}.

\bigskip

We have seen in the previous section that for the Gauss-Bonnet theory coupled
to a Maxwell field one can indeed construct black branes. In the following section
we construct new solutions for cubic and quartic Lovelock theories supported
by a three-form and four-form field strength, respectively, describing black
strings and branes.

\bigskip

\section{New charged black p-branes in Lovelock gravity}

In the previous section we have shown that the Lovelock theory of order $n$
coupled to a $(p-1)$-form with $p=n$ admits black brane solutions, provided
one can find a black object on the brane. In this section we construct new
solutions in this class for cubic and quartic Lovelock theory coupled to a
two-form and a three-form, respectively.

\subsection{Cubic Lovelock theory and a Kalb-Ramond field}

Let us propose the following ansatz for the metric:
\begin{align}
ds^{2}  &  =d\tilde{s}_{d}^{2}+d\vec{x}_{q}^{2}\label{anz1}\\
&  =-f\left(  r\right)  dt^{2}+\frac{dr^{2}}{f\left(  r\right)  }+r^{2}\left(
d\Omega_{\left(  1\right)  }^{2}+\cdots+d\Omega_{\left(  m\right)  }^{2}\right)
+d\vec{x}_{q}^{2}\ , \label{anz2}%
\end{align}
where $d\Omega_{\left(  i\right)  }$ stands for the line element of the $i$th
three-dimensional, constant curvature manifold of curvature $\gamma$, normalized
to $\pm1$ (it can be checked that the case $\gamma=0$ does not lead to black
hole solutions). For the sake of concreteness one could use the following
coordinate chart $(x_{i},y_{i},z_{i})$ on the $i$th three-dimensional
manifold with line element $d\Omega_{\left(  i\right)  }$ such that%
\begin{equation}
d\Omega_{\left(  i\right)  }^{2}=\frac{dx_{i}^{2}+dy_{i}^{2}+dz_{i}^{2}%
}{\left(  1+\frac{\gamma}{4}\left(  x_{i}^{2}+y_{i}^{2}+z_{i}^{2}\right)
\right)  ^{2}}\text{ (no sum over }i\text{)}\ ,
\end{equation}
and the following, natural, magnetic ansatz for the three-form field strength%
\begin{equation}
F_{\left(  3\right)  }=P\sum_{i=1}^{m}Vol\left(  \Omega_{\left(  i\right)
}\right)  \ ,
\end{equation}
where $P$ is a constant and which in components reads%
\begin{equation}
F_{\mu\nu\lambda}^{\left(  3\right)  }=\frac{P}{\left(  1+\frac{\gamma}%
{4}\left(  x_{1}^{2}+y_{1}^{2}+z_{1}^{2}\right)  \right)  ^{3}}\delta
_{\lbrack\mu}^{x_{1}}\delta_{\nu}^{y_{1}}\delta_{\lambda]}^{z_{1}}%
+\cdots +\frac{P}{\left(  1+\frac{\gamma}{4}\left(  x_{m}^{2}+y_{m}^{2}+z_{m}%
^{2}\right)  \right)  ^{3}}\delta_{\lbrack\mu}^{x_{m}}\delta_{\nu}^{y_{m}%
}\delta_{\lambda]}^{z_{m}}\ \text{(no sum over }m\text{)\ }.
\end{equation}

The dimension of the spacetime is therefore $D=d+q=(2+3m)+q\ $.

In this manner, as usual, the equations for the two-form are fulfilled:%
\begin{equation}
\nabla_{\mu}F_{(3)}^{\mu\nu\lambda}=0\ ,
\end{equation}
without any constraint on the magnetic charge $P$. One therefore has to deal
with the generalized Einstein equations,%
\begin{equation}
\mathcal{E}_{AB}^{\left(  3,3\right)  }:=c_{3}\mathcal{G}_{AB}^{\left(
3\right)  }-\left(  F_{ACD}F_{B}^{\ CD}-\frac{1}{6}g_{AB}F_{CDE}%
F^{CDE}\right)  =0\ ,
\end{equation}
and then solve the lapse function $f\left(  r\right)  $ from the system,%
\begin{equation}
-\frac{c_{3}}{2^{4}}\delta_{BD_{1}\cdots D_{6}}^{AC_{1}\cdots C_{6}}R_{\ \ C_{1}C_{2}%
}^{D_{1}D_{2}}R_{\ \ C_{3}C_{4}}^{D_{3}D_{4}}R_{\ \ C_{5}C_{6}}^{D_{5}D_{6}%
}-\left(  F_{\ CD}^{A}F_{B}^{\ CD}-\frac{1}{6}\delta_{B}^{A}F_{CDE}%
F^{CDE}\right)  =0\ .
\end{equation}
Note that we have explicitly included the coupling \footnote{For some explicit examples of solutions including charged black holes with planar horizons, Taub-Nut black holes and their coupling with nonlinear electrodynamics in general Lovelock theories containing the cubic term see the works in Ref. \cite{DHWork}.} $c_{3}$. 

With this ansatz, one can use the results from the previous sections to show
that the equations $\mathcal{E}_{x_{i}x_{i}}^{\left(  3,3\right)  }=0$ (no sum
over $x_{i}$) are implied by the trace of the equations on the brane. Out of
these equations one can show that%
\begin{equation}
\mathcal{E}_{t}^{\left(  3,3\right)  t}=\mathcal{E}_{r}^{\left(  3,3\right)
r}\ .
\end{equation}
The equations projected along the angles $\Omega_{\left(  i\right)  }$ are
all equal and reduce to a linear combination of the $\mathcal{E}_{t}^{\left(
3,3\right)  t}$ equation and its derivative with respect to the radial
coordinate, as expected from diffeomorphism invariance. Consequently, one is
left with a single master equation that admits a first integral and leads to
the following Wheeler's-like polynomial for $f=f\left(  r\right)  $:%
\begin{equation}
f^{3}-\frac{6\gamma}{(d-3)}f^{2}+\frac{12}{(d-4)(d-3)}f+\frac{1}{54}%
\frac{(d-8)!P^{2}}{(d-3)!c_{3}}+\frac{\mu}{r^{d-7}c_{3}}-\frac{8\gamma
(d-8)}{(d-6)(d-7)(d-3)(d-4)}=0\ . \label{WP33}%
\end{equation}
Here $\mu$ is an integration constant.

Assuming the existence of a horizon located at $r=r_{+}>0$, we can compute the
temperature and the entropy density of the black string \cite{JACOBSONMYERS},
which respectively read%
\begin{align}
T  &  =\frac{f^{\prime}\left(  r_{+}\right)  }{4\pi}=\frac{1}{48}%
\frac{(d-7)(d-4)(d-3)\mu}{c_{3}\pi r_{+}^{d-6}}\ ,\\
s  &  =64\pi^{2}c_{3}(d-5)(d-2)r_{+}^{d-6},
\end{align}
and $\mu=\mu\left(  r_{+}\right)  $ is given by%
\begin{equation}
\mu\left(  r_{+}\right)  =\frac{\left(  d-2\right)  !\left(  d-5\right)
}{\left(  d-7\right)  !}\left(  8\gamma c_{3}(d-8)-\frac{1}{54}\frac{P^{2}%
}{(d-5)}\right)  r_{+}^{d-7}\ . \label{mudermass3}%
\end{equation}
As expected, the positivity of the temperature and entropy imply $c_{3}>0$ and
$\mu>0$, which by virtue of (\ref{mudermass3}) and for $d>8$ imply $\gamma=1$
(the case $d=8$ does not lead to black holes). Therefore, the horizon of the
black branes is given by the product of $m$ three-spheres times $\mathrm{I\!R}%
^{q}$. Figure 1 depicts the lapse function for some values of the integration
constants in dimensions $d=11$ and $d=14$.

\begin{figure}
     \includegraphics[scale=0.5]{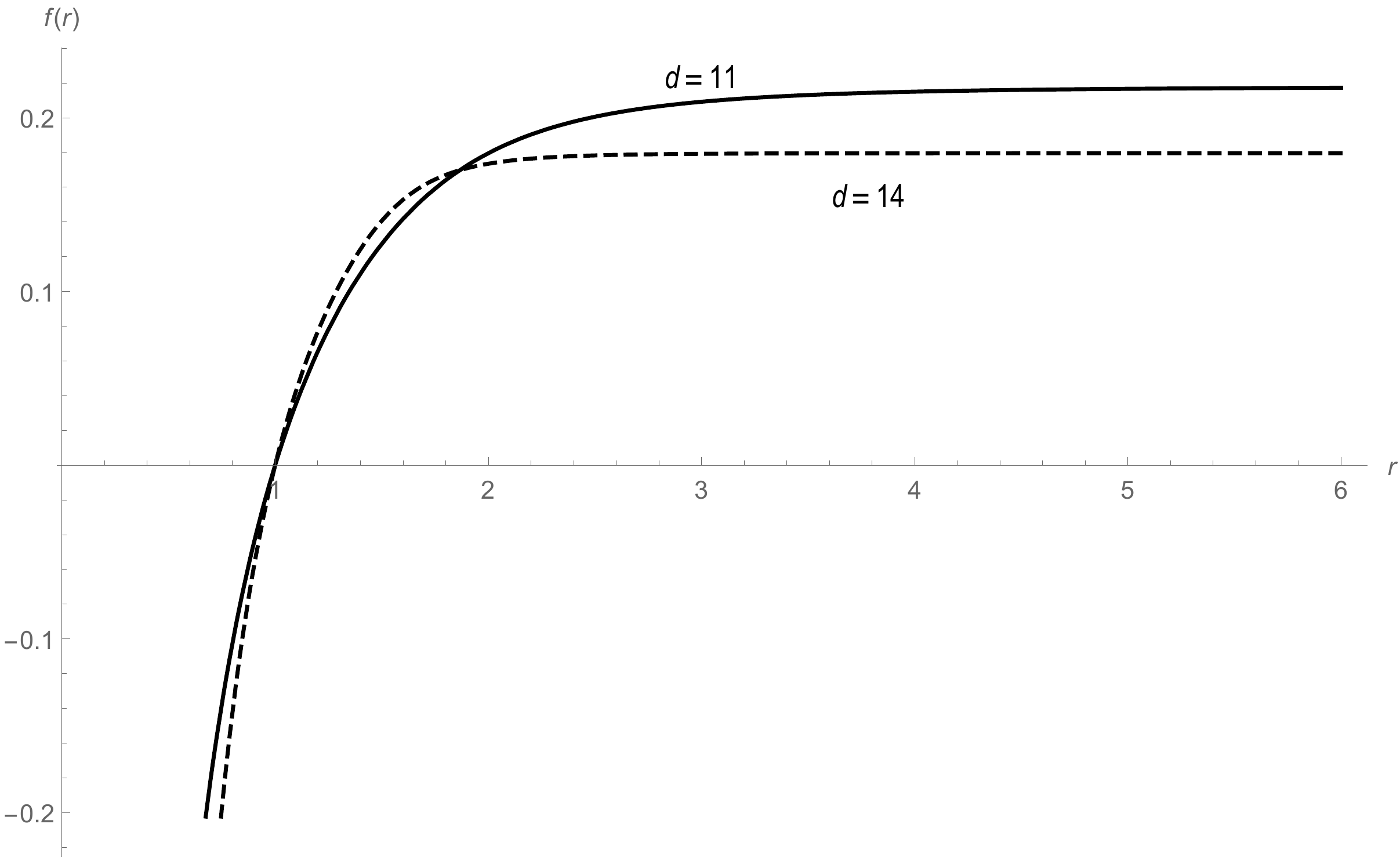} \caption{\textit{Charged black holes for cubic Lovelock theory supported by a three-form field strength. $f(r)$ for $c_3=1$, $r_+=1$ and magnetic charge $P=1$, for dimensions $d=11$ and $d=14$. The event horizon is a product of three-spheres and this solution can be trivially oxidated to higher dimensions by adding a factor $\mathrm{I\!R}^q$ to the metric.}}  
\end{figure}

The solution with vanishing integration constants, $\mu=P=0$, has the form%
\begin{equation}
ds_{0}^{2}=-\alpha_{d}\left(  P=0,\mu=0\right)  dt^{2}+\frac{dr^{2}}{\alpha
_{d}\left(  P=0,\mu=0\right)  }+r^{2}\left(  d\Omega_{\left(  1\right)  }%
^{2}+\cdots +d\Omega_{\left(  m\right)  }^{2}\right)  +d\vec{x}_{q}^{2}\ ,
\end{equation}
with $\alpha_{d}$ a real solution of the cubic polynomial (\ref{WP33}). This
spacetime has a curvature singularity at the origin that, as shown in Fig. 1, can be covered by an
event horizon for nonvanishing values of the integration constants. The
situation is similar to what occurs for Lifsthiz spacetimes, where the
Lifshitz background is singular at the origin (it contains timelike incomplete geodesics), but this singularity is hidden
by an event horizon in the case of Lifshitz black holes (see e.g. \cite{Lif}).
This behavior is also reminiscent of gravitating monopoles \cite{GravMon}.

The asymptotic behavior of the lapse function is%
\begin{equation}
f\left(  r\right)  =\alpha_{d}\left(  P\right)  -\frac{\tilde{\mu}\left(
d,\mu,P\right)  }{r^{\frac{d-7}{3}}}+\cdots\ ,
\end{equation}
where $\cdots$ denotes subleading terms. Here $\alpha_d$ and $\tilde{\mu}$ depend
on the dimension as well as on the integration constants $\mu$ and $P$. It is
interesting to notice that the first dependence in $r$ has the usual falloff
for the black hole solution of the cubic Lovelock theory in vacuum \cite{BHS}, therefore one
may expect that for a fixed $P=\bar{P}$, these solutions should have a finite
mass when the energy density is measured with respect to the background%
\begin{equation}
ds_{\infty}^{2}=-\alpha_{d}\left(  \bar{P}\right)  dt^{2}+\frac{dr^{2}}%
{\alpha_{d}\left(  \bar{P}\right)  }+r^{2}\left(  d\Omega_{\left(  1\right)
}^{2}+\cdots +d\Omega_{\left(  m\right)  }^{2}\right)  +d\vec{x}_{q}%
^{2}\ ,\label{asymp}%
\end{equation}
with a nonvanishing magnetic flux. It is interesting to notice also that the
term that decays as $r^{-(d-7)/3}$ has a slower falloff than that of an
asymptotically flat solution with finite mass in general relativity $(r^{-(d-3)})$, nevertheless since only the cubic term is present, the mass will
come from the term with falloff $r^{-(d-7)/3}$ on the lapse function.

\bigskip

\subsection{Quartic Lovelock theory and a three-form: Magnetically charged
solutions}

Let us now construct new black brane solutions of quartic Lovelock theory
coupled to a four-form field strength. We will first construct solutions with
magnetic charge only, for which we use the ansatz (\ref{anz1}) and (\ref{anz2}),
where now $d\Omega_{\left(  i\right)  }$ stands for the line element of the
$i$th four-dimensional, constant curvature manifold of curvature $\gamma$,
normalized to $\pm1$. Again for concreteness a coordinate chart $(x_{i}%
,y_{i},z_{i},w_{i})$ on the $i$th four-dimensional manifold with line element
$d\Omega_{\left(  i\right)  }$ can be used and then%
\begin{equation}
d\Omega_{\left(  i\right)  }^{2}=\frac{dx_{i}^{2}+dy_{i}^{2}+dz_{i}^{2}%
+dw_{i}^{2}}{\left(  1+\frac{\gamma}{4}\left(  x_{i}^{2}+y_{i}^{2}+z_{i}%
^{2}+w_{i}^{2}\right)  \right)  ^{2}}\text{ (no sum over }i\text{)}\ ,
\end{equation}
and now the magnetic ansatz for the four-form reads%
\begin{equation}
F_{\left(  4\right)  }=P\sum_{i=1}^{m}Vol\left(  \Omega_{\left(  i\right)
}\right)  \ ,
\end{equation}
which in components turn out to be%
\begin{equation}
F_{\mu\nu\lambda\rho}^{\left(  4\right)  }=\frac{P}{\left(  1+\frac{\gamma}%
{4}\left(  x_{1}^{2}+y_{1}^{2}+z_{1}^{2}+w_{1}^{2}\right)  \right)  ^{3}%
}\delta_{\lbrack\mu}^{x_{1}}\delta_{\nu}^{y_{1}}\delta_{\lambda}^{z_{1}}%
\delta_{\rho]}^{w_{1}}+\cdots +\frac{P}{\left(  1+\frac{\gamma}{4}\left(
x_{m}^{2}+y_{m}^{2}+z_{m}^{2}+w_{m}^{2}\right)  \right)  ^{3}}\delta
_{\lbrack\mu}^{x_{m}}\delta_{\nu}^{y_{m}}\delta_{\lambda}^{z_{m}}\delta
_{\rho]}^{w_{1}}\text{\ },
\end{equation}
with no sum over $m$. Now the spacetime dimension is $D=d+q=(2+4m)+q\ $.
Again, Maxwell equations for the fundamental three-form are trivially
fulfilled and introducing the gravitational coupling $c_{4}$ one is left with
the system%
\begin{equation}
\mathcal{E}_{AB}^{\left(  4,4\right)  }:=c_{4}\mathcal{G}_{AB}^{\left(
4\right)  }-\left(  F_{ACDE}F_{B}^{\ CDE}-\frac{1}{8}g_{AB}F_{CDEH}%
F^{CDEH}\right)  =0\ \text{,}%
\end{equation}
which explicitly reads%
\begin{equation}
-\frac{c_{4}}{2^{5}}\delta_{BD_{1}\cdots D_{8}}^{AC_{1}\cdots C_{8}}R_{\ \ C_{1}C_{2}%
}^{D_{1}D_{2}}R_{\ \ C_{3}C_{4}}^{D_{3}D_{4}}R_{\ \ C_{5}C_{6}}^{D_{5}D_{6}%
}R_{\ \ C_{7}C_{8}}^{D_{7}D_{8}}-\left(  F_{\ CDE}^{A}F_{B}^{\ CDE}-\frac
{1}{8}g_{AB}F_{CDEH}F^{CDEH}\right)  =0\ .\label{quartic}%
\end{equation}

As before, since there is a single metric function and the system comes from
an action that is invariant under diffeomorphisms, it is enough to integrate
the ($tt$) equation that leads to the following Wheeler's-like quartic polynomial:%

\begin{align*}
f^{4}-\frac{12\gamma}{(d-3)}f^{3} &  +\frac{18(3d-16)\gamma^{2}}%
{(d-5)(d-4)(d-3)}f^{2}-\frac{108\gamma(d-8)}{(d-7)(d-5)(d-4)(d-3)}f\\
-\frac{1}{384}\frac{(d-10)!P^{2}}{(d-3)!c_{4}} &  +\frac{27(3d^{2}%
-48d+188)}{(d-9)(d-8)(d-7)(d-5)(d-4)(d-3)}-\frac{\mu}{r^{d-9}c_{4}}=0\ .
\end{align*}

Again, $\mu$ is an integration constant. The expression for the temperature and
entropy density of a black hole with horizon located at $r=r_{+}$ read%
\begin{align}
T &  =\frac{f^{\prime}\left(  r_{+}\right)  }{4\pi}=\frac{1}{432}\frac
{\mu(d-9)(d-7)(d-5)(d-4)(d-3)}{(d-8)\gamma\pi c_{4}r_{+}^{d-8}}\ ,\\
s &  =216\pi^{3}\gamma(d-8)(d-6)(d-2)c_{4}r_{+}^{d-8}\ ,
\end{align}
and in this case $\mu=\mu\left(  r_{+}\right)  $ is given by%

\begin{equation}
\mu\left(  r_{+}\right)  =\frac{(d-10)!}{(d-3)!}\left[  27(3d^{2}%
-48d+188)(d-6)c_{4}-\frac{1}{384}P^{2}\right]  r_{+}^{d-9}\ .
\end{equation}

Considering the latter expression plus restricting the entropy and temperature
to be positive, implies $c_{4}>0$, $\mu>0$ and $\gamma=1$. Consequently, in
this case the event horizon of the black branes has the local geometry of the
product of $m$ four-spheres times $\mathrm{I\!R}^{q}$. In Fig. 2 we show
some plots of the lapse functions for different values of the integration
constants and dimensions.

\begin{figure}%
    \subfloat[]{{\includegraphics[scale=0.3]{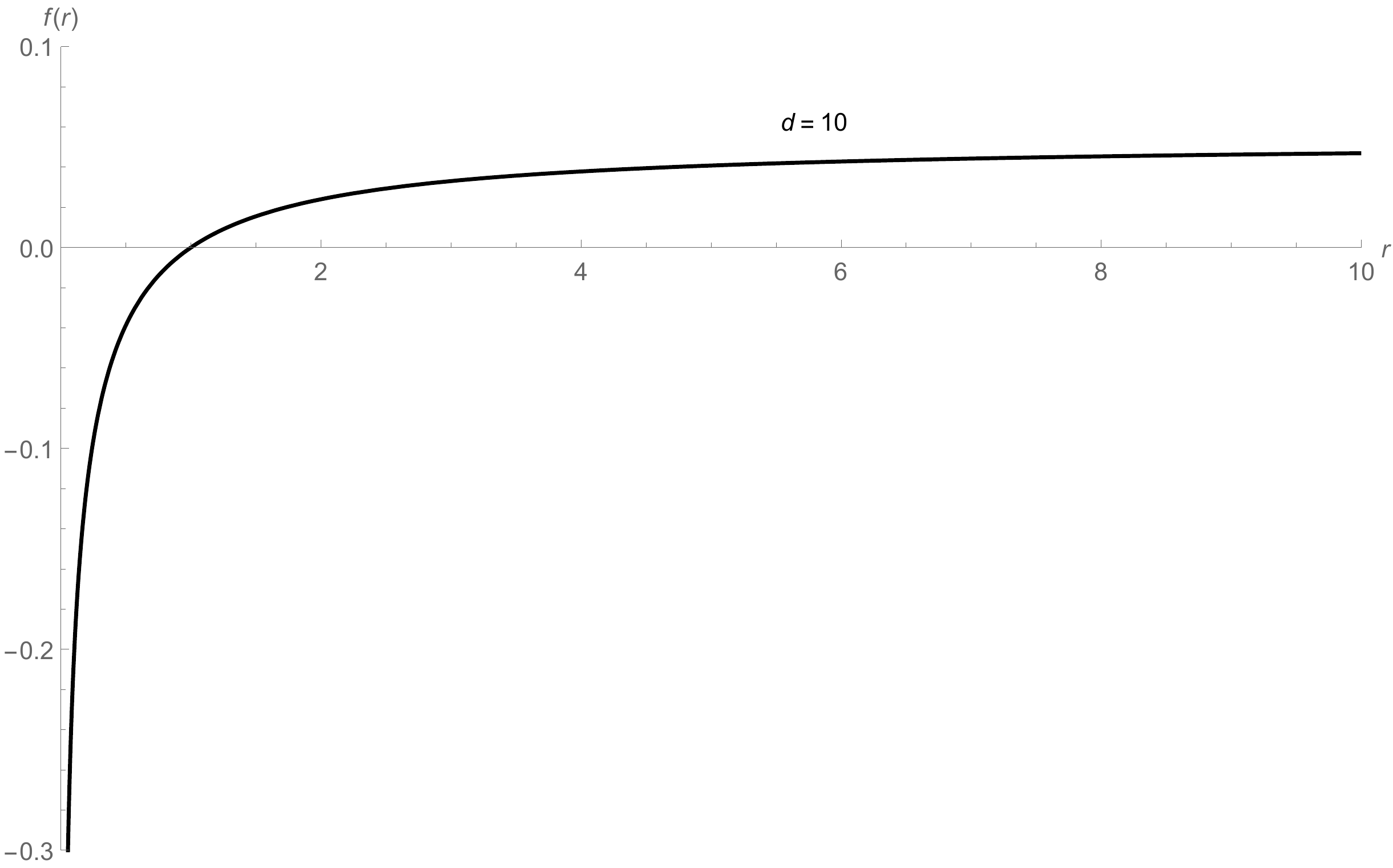} }}%
    \qquad
    \subfloat[]{{\includegraphics[scale=0.3]{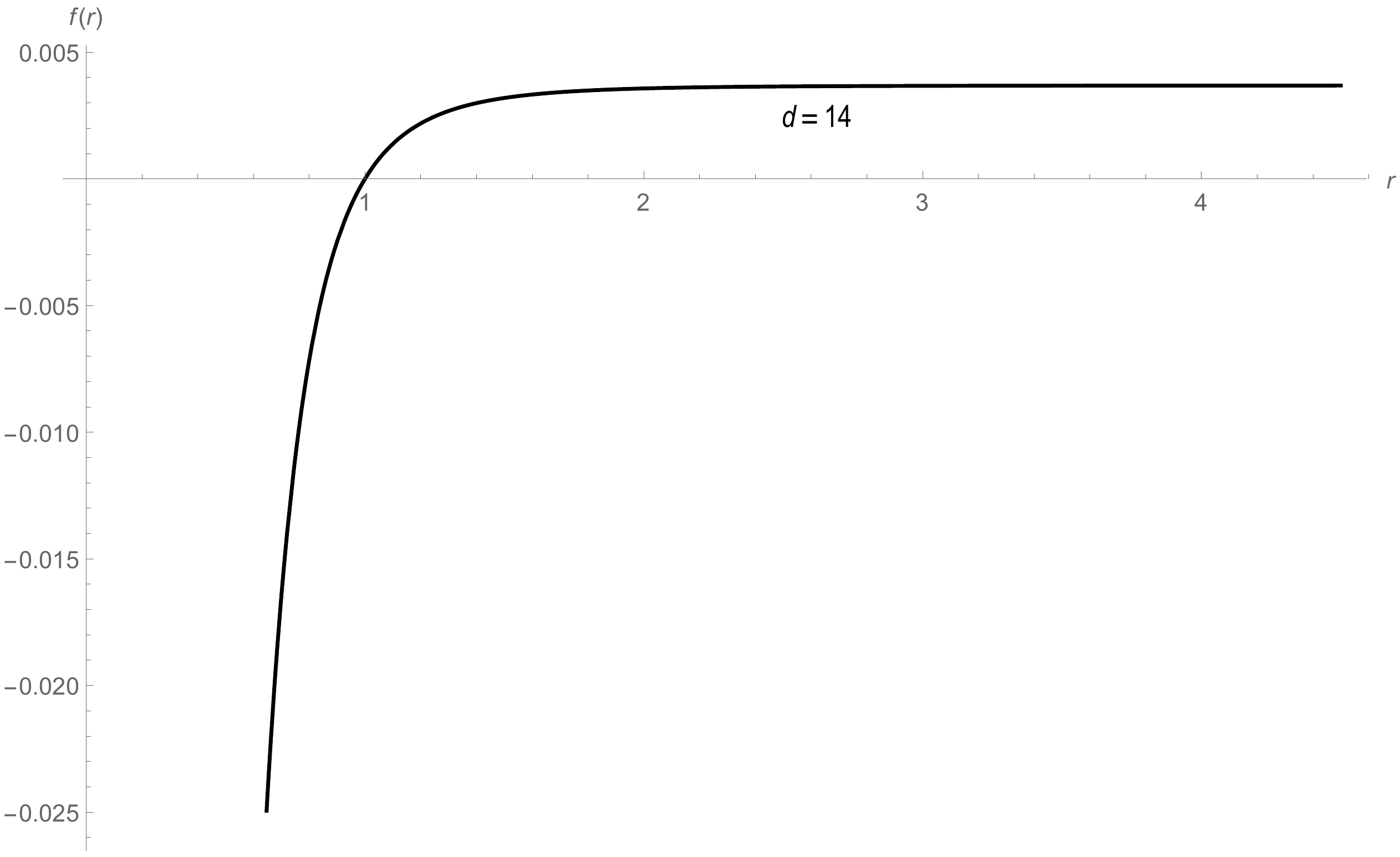} }}%
    \caption{\textit{Magnetically charged black holes for quartic theory supported by a four-form field strength in dimensions $d=10$ (left panel with $P=2900$) and $d=14$ (right panel with $P=550$). In both figures we have chosen $c_4=1$ and $r_+=1$. The event horizon is given by the product of four-spheres and these solutions can be cylindrically oxidated to higher dimensions by adding a factor $\mathrm{I\!R}^q$ to the metric.}}
    \label{quartics}%
\end{figure}

As before, when $r\rightarrow +\infty$, the lapse function approaches a constant
$\alpha_{d}\left(  P\right)  $ plus a term proportional to a function
$\tilde{\mu}\left(  d,\mu,P\right)  $ that decays as $r^{-\left(  d-9\right)
/3}$. This coincides with the asymptotic falloff of the solution in quartic Lovelock theory in vacuum
\cite{BHS}.

\subsection{Quartic Lovelock theory and a three-form: Dyonic solutions}

Finally, let us construct new black brane solutions with both electric and
magnetic charge, i.e. dyonic solutions. We assume that the line element is
given by (\ref{anz1}) and (\ref{anz2}) and now $d\Omega_{\left(  i\right)  }$
stands for the line element of the $i$th two-dimensional manifold of constant
curvature $\gamma$, normalized to $\pm1$. One can choose the coordinate patch
$(x_{i},y_{i})$ on the $i$th two-dimensional manifold such that%
\begin{equation}
d\Omega_{\left(  i\right)  }^{2}=\frac{dx_{i}^{2}+dy_{i}^{2}}{\left(
1+\frac{\gamma}{4}\left(  x_{i}^{2}+y_{i}^{2}\right)  \right)  ^{2}}\text{ (no
sum over }i\text{)}\ .
\end{equation}
In order to turn on the electric and magnetic parts of the field strength we
assume%
\begin{equation}
F_{\left(  4\right)  }=Qdt\wedge dr\wedge\sum_{i}Vol\left(  \Omega_{\left(
i\right)  }\right)  +P\sum_{i<j}Vol\left(  \Omega_{\left(  i\right)  }\right)
\wedge Vol\left(  \Omega_{\left(  j\right)  }\right)  \ ,
\end{equation}
which in components reads%
\begin{equation}
F_{\mu\nu\lambda\rho}^{\left(  4\right)  }=P\sum_{i<j}^{m}\frac{\delta
_{\lbrack\mu}^{x_{i}}\delta_{\nu}^{y_{i}}\delta_{\lambda}^{x_{j}}\delta
_{\rho]}^{y_{j}}}{\left(  1+\frac{\gamma}{4}\left(  x_{i}^{2}+y_{i}%
^{2}\right)  \right)  \left(  1+\frac{\gamma}{4}\left(  x_{j}^{2}+y_{j}%
^{2}\right)  \right)  }+Q\sum_{i}^{m}\frac{\delta_{\lbrack\mu}^{t}%
\delta_{\nu}^{r}\delta_{\lambda}^{x_{i}}\delta_{\rho]}^{y_{i}}}{\left(
1+\frac{\gamma}{4}\left(  x_{i}^{2}+y_{i}^{2}\right)  \right)  },
\end{equation}
The spacetime dimension is now $D=d+q=(2+2m)+q\ $. Maxwell equations for the
fundamental three-form are fulfilled identically and as before the system
(\ref{quartic}) leads to a single polynomial equation that resembles Wheeler's
polynomial,
\begin{align}
&  f^{4}-\frac{4\gamma}{(d-3)}f^{3}+\frac{6\gamma^{2}}{(d-5)(d-3)}f^{2}%
-\frac{4\gamma}{(d-7)(d-5)(d-3)}f-\frac{\mu}{r^{d-9}c_{4}}\nonumber\\
&  -\frac{1}{768}\frac{(d-10)!(d-4)P^{2}}{c_{4}(d-3)!}+\frac{1}{192}%
\frac{(d-9)!Q^{2}}{c_{4}(d-7)(d-3)!r^{2(d-8)}}+\frac{\gamma^{2}}%
{(d-9)(d-7)(d-5)(d-3)}=0\ .
\end{align}
Temperature and entropy density read%
\begin{align}
T &  =\frac{f^{\prime}\left(  r_{+}\right)  }{4\pi}=\frac{(d-5)(d-3)}%
{16\gamma\pi c_{4}r_{+}^{d-8}}\left[  \mu(d-9)(d-7)-\frac{1}{96}%
\frac{(d-9)!(d-8)}{(d-3)!}\frac{Q^{2}}{r_{+}^{d-7}}\right]  \ ,\\
s &  =16\pi^{2}\gamma c_{4}(d-6)(d-4)(d-2)r_{+}^{d-8}\ ,
\end{align}
and $\mu=\mu\left(  r_{+}\right)  $ in this case is given by

\[
\mu\left(  r_{+}\right)  =\frac{r_{+}^{d-9}}{(d-7)(d-5)(d-3)}\left(
\frac{\gamma^{2}c_{4}}{(d-9)}+\frac{1}{192(d-8)(d-6)}\left[  \frac
{Q^{2}r_{+}^{2(8-d)}}{(d-7)(d-4)}-\frac{P^{2}}{4(d-9)}\right]  \right)  \ .
\]

Figure 3 contains the lapse functions of dyonic black branes in quartic
Lovelock theory with horizons given by the product of $m$ two-spheres times
$\mathrm{I\!R}^{q}$.

\begin{figure}
     \includegraphics[scale=0.5]{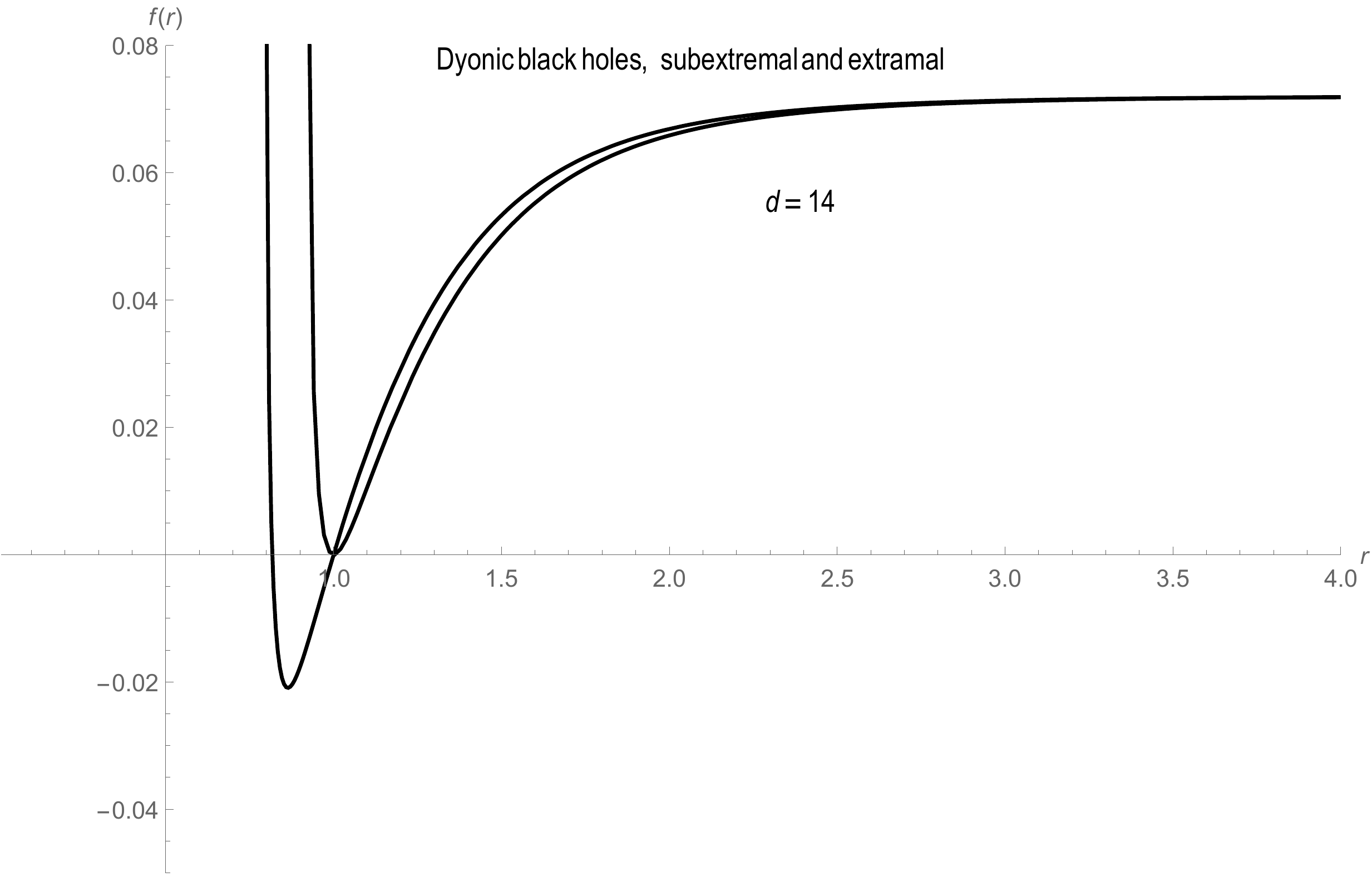} \caption{\textit{Dyonic solution in quartic Lovelock theory in dimension $d=14$. The event horizon is given by a product of six two-spheres and we have chosen $c_4=1$, $P=1$ and $r_+=1$. Two different values for the electric charge are presented which correspond to a subextremal value that leads to a spacetime with an event and Cauchy horizon and also the extremal case for which the horizons degenerate and the solution has vanishing temperature.}}  
\end{figure}

\section{Conclusions and further comments}

In this paper we have shown that Lovelock theories with a single Lovelock term\footnote{Lovelock theories with a single Lovelock term also admit for black holes with nonminimally coupled scalar hairs \cite{HC}.} 
of order $n$ admit cylindrical, homogeneous charged black strings in the presence of a
($p-1$)-form, only when $p=n$. In the simplest case, one can consider general
relativity ($n=1$) coupled to a massless scalar field ($p-1=0$) and construct
branelike solutions by just adding flat directions. No-hair theorems in this
case forbid the existence of black objects with a nonvanishing scalar field.
We have explicitly shown that the next simpler case, i.e. Gauss-Bonnet-Maxwell
theory, does admit charged black strings and black branes which are homogeneous
on the extended directions. We have extended these results and constructed new explicit solutions, for the cubic Lovelock theory supported by a
Kalb-Ramond field as well as for the quartic Lovelock theory with a four-form
field strength.

The new black brane solutions we have constructed here have a horizon geometry
that is given by the product of spheres times
$\mathrm{I\!R}^{q}$. It is well known that in general relativity in vacuum the
usual sphere of Schwarzschild-Tangherlini solution \cite{Tangherlini:1963bw},
hereafter the base manifold, can be replaced by an arbitrary Euclidean
Einstein manifold \cite{Gibbons:2002pq}. In Einstein-Gauss-Bonnet theory in
vacuum, it was realized by Dotti and Gleiser \cite{Dotti:2005rc} that for
generic values of the Gauss-Bonnet coupling, an Einstein base manifold must
also fulfill a quadratic tensor constraint on the curvature that admits a
simple expression when written in terms of the Weyl tensor. Some explicit
examples of horizons were given in \cite{Maeda:2010bu} as well as the
expression for the finite quasilocal mass. In Refs. \cite{DOT5},
\cite{DOT6}, \cite{DOT7}, it was shown that for generic values of the
couplings, the Einstein restriction on the base manifold is a necessary
condition and that for the particular relation that leads to a theory with a
single maximally symmetric solution, the restrictions are weakened and the set
of possible causal structures is enlarged to admit as well wormhole solutions
in vacuum \cite{WORMHOLE}. This analysis was extended to Lovelock theories in
the Chern-Simons case \cite{Zanelli:2005sa} in Ref. \cite{Oliva:2012ff}
obtaining similar results even with time dependence. Also products of Thurston
geometries may appear as models for the horizon in Lovelock theories with a
unique maximally symmetric solution \cite{Anabalon:2011bw}. For arbitrary
values of the Lovelock couplings a thorough analysis was provided in Ref.
\cite{Ray:2015ava}, where it was shown that the base manifold has to fulfill a
hierarchy of Euclidean Lovelock equations. A similar result was recently
obtained for warped black strings in \cite{Kastor:2017knv}.

\bigskip

Of particular
relevance for the present work is Ref. \cite{Bardoux:2012aw}, where new
static solutions in general relativity coupled to $p$-forms were presented.
Here we have used the same ansatz for the $p$-forms in the electric and
magnetic cases, where the components are distributed on a symmetric fashion
along the base manifold. These results were extended to the
Einstein-Gauss-Bonnet theory in \cite{Bardoux:2010sq}. In the latter
references the authors also consider base manifolds that are products of
Einstein-K\"{a}hler spaces, with a $p$-form field strength that is
proportional to exterior products of the K\"{a}hler forms of each manifold.
An exhaustive analysis of the solutions supported by a $p$-form in general
relativity in the Robinson-Trautman family was provided in
\cite{Ortaggio:2014gma} and \cite{Ortaggio:2016zuk}. All these solutions, as
the ones presented in Sec. IV of this paper, have an asymptotic behavior
that matches the transverse section of the metric (\ref{asymp}). These metrics
have a singularity at $r=0$, which for black hole spacetimes is surrounded by
an event horizon.

\bigskip

The stability of homogeneous black strings and black branes is easier to study
than that of the nonhomogeneous objects\footnote{Even though the Cauchy problem for theories containing a unique Lovelock term could be ill posed around a generic background, Refs. \cite{US1} and \cite{US2} show that the linearized problem can be explored in a systematic fashion, providing a well-defined result. As it occurs in vacuum for spherically symmetric black holes, we believe that the configurations we have constructed in this work will appear as the high-curvature regime of charged black string solutions including also the Einstein-Hilbert term. Those solutions might be constructed numerically and in such a setup one could hope to study the nonlinear evolution of the perturbation.}. Since the solutions are symmetric
under translations along the extended directions, one can construct plane
waves with a definite momentum along such directions. In general relativity
\cite{GL}, Gauss-Bonnet \cite{US1} and cubic Lovelock theories \cite{US2}
there is a perturbative instability, a Gregory-Laflamme instability, that it
is triggered by modes with wavelength above a critical value, and it is
therefore a long wavelength instability, sharing some properties of
instabilities of other physical systems \cite{Cardoso:2006ks}. It would be
interesting to explore the stability of the solutions here constructed, in particular that of the extremal dyonic case in quartic Lovelock theory.

\bigskip

\section*{Acknowledgements}

M.L. and A.V. appreciate the support of CONICYT Fellowships No. 21141229 and No. 21151067, respectively. J. O. thanks Gast\'{o}n Giribet for enlightening
discussions. This work was also funded by CONICYT Grant No. DPI20140053 (J.O. and
A.G.) and FONDECYT Grants No. 1150246 and No. 1181047.

\end{document}